Original article

# Local dynamic stability of treadmill walking: intrasession and week-to-week repeatability

Authors:

Fabienne Reynard[ab], PT MSc

Philippe Terrier[ab§], PhD

[a] IRR, Institute for Research in Rehabilitation, Sion, Switzerland

[b] Clinique romande de réadaptation SUVACare, Sion, Switzerland

**§ Corresponding author:**

Dr Philippe Terrier

Clinique romande de réadaptation SUVACare

Av. Gd-Champsec 90

1951 Sion

Switzerland

Tel.: +41-27-603-23-91

E-mail: Philippe.Terrier@crr-suva.ch

**Abstract**

Repetitive falls degrade the quality of life of elderly people and of patients suffering of various neurological disorders. In order to prevent falls while walking, one should rely on relevant early indicators of impaired dynamic balance. The local dynamic stability (LDS) represents the sensitivity of gait to small perturbations: divergence exponents (maximal Lyapunov exponents) assess how fast a dynamical system diverges from neighbor points. Although numerous findings attest the validity of LDS as a fall risk index, reliability results are still sparse. The present study explores the intrasession and intersession repeatability of gait LDS using intraclass correlation coefficients (ICC) and standard error of measurement (SEM). Ninety-five healthy individuals performed 5min. treadmill walking in two sessions separated by 9 days. Trunk acceleration was measured with a 3D accelerometer. Three time scales were used to estimate LDS: over 4 to 10 strides ($\lambda_{4-10}$), over one stride ($\lambda_1$) and over one step ($\lambda_{0.5}$). The intrasession repeatability was assessed from three repetitions of either 35 strides or 70 strides taken within the 5min tests. The intersession repeatability compared the two sessions, which totalized 210 strides. The intrasession ICCs (70-strides estimates/35-strides estimates) were 0.52/0.18 for $\lambda_{4-10}$ and 0.84/0.77 for $\lambda_1$ and $\lambda_{0.5}$. The intersession ICCs were around 0.60. The SEM results revealed that $\lambda_{0.5}$ measured in medio-lateral direction exhibited the best reliability, sufficient to detect moderate changes at individual level (20%). However, due to the low intersession repeatability, one should average several measurements taken on different days in order to better approximate the true LDS.

# 1. Introduction

Falls are a major health issue in older adults (Stevens et al., 2006). Moreover, repetitive falls degrade the quality of life of patients suffering of various neurological disorders (Finlayson et al., 2006; Kerr et al., 2010; Ramnemark et al., 2000). It has been shown that stride-to-stride variability of gait kinematics is related to fall risk (Brach et al., 2007; Verghese et al., 2009). This increased variability may be the result of a decreased ability to optimally control gait. Several analytical methods have been developed to better take into account the nonlinear features of gait variability (Hamacher et al., 2011; Hausdorff, 2007; Stergiou and Decker, 2011). Originally developed to detect deterministic chaos in nonlinear dynamical systems, the maximal Lyapunov exponent has been advocated as a relevant method to assess the degree of sensitivity of gait to small perturbations, or in other words the local dynamic stability (LDS). Computed from various continuously measured kinematic parameters (speed, acceleration, joint angles), the LDS represents the rate of divergence between neighbor trajectories in a reconstructed state space that describes the dynamics of the system (Dingwell, 2006; Dingwell and Cusumano, 2000; Terrier and Dériaz, 2013).

Some recent experimental and clinical findings support the hypothesis that LDS could be used to predict fall risk (Bruijn et al., 2011; Lockhart and Liu, 2008; McAndrew et al., 2011; Roos and Dingwell, 2010; Toebes et al., 2012). The validity of LDS compared to other indicators has been recently discussed and LDS was found to be one of the best stability indices (Bruijn et al., 2013). However, several issues must be solved before LDS can be routinely used as an early fall-risk predictor in clinical settings. In the first place, information regarding the reliability of LDS measurements is still sparse (Bruijn et al., 2010a; Kang and Dingwell, 2006). Recently, a good intrasession repeatability (ICC between 0.79 and 0.92) has been described by analyzing long duration outdoor walking (200 strides) in young healthy subjects (N=20) (van Schooten et al., 2013). The same study reports lower intersession (between nonconsecutive days) repeatability lying between 0.38 and 0.63. However, more results are needed about the LDS reliability in short indoor walking tests, which are more adapted to patients with diminished walking capabilities (Hilfiker et al., 2013; Terrier et al., 2013).

Other research questions need to be addressed in order to increase the usability of LDS in clinical settings. Two different methods are used to characterize short-term LDS, which is assumed to be the more relevant time scale for assessing gait stability (Bruijn et al., 2011; Roos and Dingwell, 2010): one method computes divergence over the duration of one stride (Manor et al., 2009; McAndrew Young and Dingwell, 2012), and the other one over the duration of one step (Bruijn et al., 2009; Toebes et al., 2012). It is still not known whether one method offers more precision than the other one. In addition, it is known that a minimal length of continuous signal is necessary to assess the maximal Lyapunov exponents (Kang and Dingwell, 2006; Rosenstein et al., 1993). However, the averaging of several estimates of LDS obtained from distinct short-duration signals produces reliable results, at least at group level (Sloot et al., 2011). Thereby two possibilities should be distinguished: 1) Inherent to the calculation method, a too short-duration signal induces very large error on LDS estimates: long-duration walking tests are therefore mandatory; 2) LDS can be precisely assessed from a short-duration signal, but it substantially varies from strides to strides: as a result, the precision of estimate can be increased by averaging the results obtained from several short-duration walking tests.

To address the aforementioned issues, the LDS of 95 healthy adults walking on a treadmill was assessed from trunk acceleration signals. We aimed to answer the following research questions: 1) What is the intrasession (within a 5min. continuous measurement) and intersession (week-to-week) repeatability (i.e. absolute agreement among repetitions) of LDS? 2) How reliable is an estimate of LDS obtained from short-duration walking tests? 3) Does the repeatability of LDS increase with measurement length as expected? 4) Do the two methods that assess short-term LDS exhibit the same repeatability?

## 2. Methods

### 2.1 Subjects

One hundred healthy subjects (50 males, 50 females) were recruited to participate in the study. They were selected according to their age and sex. Ten males and 10 females for each decade, between 20 and 69 years old, were included. The data of five subjects had to be discarded due to technical issues (age of the discarded participants (yr.): 58, 41, 22, 23, 30; 2 males, 3 females). Therefore, the final sample contained 95 participants (48 males, 47 females) whose characteristics were (mean (SD)): age 44 yr (14), body weight 70 kg (14), body height 1.72 m (0.07)). None had a history of neurological or orthopedic conditions likely to affect their mobility. Each subject gave his/her written informed consent. The study was approved by the regional medical ethics committee (Commission Cantonale Valaisanne d'Ethique Médicale, Sion, Switzerland).

### 2.2. Experimental Procedure

The subjects were equipped with a tri-axial accelerometer (Physilog system, BioAGM, Switzerland; sampling rate 200Hz, 16-bit resolution) fixed to a belt at the anterior upper trunk level under the sternal notch. The accelerometer was connected to a lightweight data logger and it measured the body accelerations along three axes: medio-lateral (ML), vertical (V) and antero-posterior (AP).

For safety reasons, subjects walked on a level treadmill (HP Cosmos, Venus model, Traunstein, Germany) wearing a harness, which didn't perturb the arms and legs movements. To remove any influence of the footwear on gait characteristics, the participants walked barefoot. The subjects walked during five minutes at preferred walking speed (PWS), assessed as described in (Dingwell and Marin, 2006). The same procedure was repeated on average $9.1 \pm 3.6$ days later. The same treadmill speed as in the first session was imposed to the participants. The data analysis was then performed with Matlab (Mathworks, MA, USA).

*2.3 Data analysis*

Because the switching on of the accelerometer might have caused transient gait perturbations, the first five seconds of the raw acceleration signals were discarded. In order to lower the effect of sensor placement variability among subjects, the 3D-acceleration signals were first re-oriented according to the procedure introduced by Moe-Nilssen (Moe-Nilssen, 1998), which uses the accelerometer's capacity as an inclinometer. The raw 200Hz signals were filtered down to 50Hz in order to accelerate the subsequent steps of data analysis; (Chebyshev Type I, zero-phase filtering). The resulting 50Hz signals (295 s duration) were divided into 3 parts of 95 s duration, beginning respectively at 5 s, 105 s and 205 s. In each of those parts, Step Frequency (SF) was assessed by using Fast Fourier Transform of the vertical acceleration signal. Then the duration corresponding to either 35 strides (1/SF x 35) or 70 strides (1/SF x 70) were taken from the beginning of each part. The resulting segments, whose length depended upon SF of each participant, were time-normalized to a uniform length of 2,500 samples (35 strides) or 5,000 samples (70 strides), using a polyphase filter implementation (matlab command *resample*). The length (2500 or 5000) corresponded approximately to the length of the longest original sample. The figure 1 presents an overall scheme, which summarizes the data pre-processing.

The method for quantifying the local dynamical stability (LDS) of gait was based on the computation of the largest Lyapunov exponent using the Rosenstein's algorithm (Dingwell, 2006; Rosenstein et al., 1993; Terrier and Dériaz, 2011). We recently published a thorough description of the method (Terrier and Dériaz, 2013). Therefore, we focus on the particularity of the methodology. A uniform time delay (AP: 5, V: 5 ML: 6) was used for all tests according to the average results of an average mutual information (AMI) analysis. A constant dimension of 6 was selected, according to the average results of a global false nearest neighbors (GFNN) analysis. Three different time scales were used to compute the slope from the logarithmic divergence curves: 1) Long-term LDS, over 4 to 10 strides ($\lambda_{4-10}$), 2) short term LDS over one stride ($\lambda_1$), 3) short-term LDS over one step ($\lambda_{0.5}$). A constant number of 71 samples per stride was used (i.e. 5000 (2500) samples for 70 (35) strides).

*2.4 Statistics*

Descriptive statistics of the basic gait parameters (speed, cadence, step length) are presented in the supplemental (online) material. In order to illustrate the difference in LDS induced by the length of the measurement, the descriptive statistics (mean, SD, median, quartiles) for 35 strides and 70 strides are presented separately in fig. 2, based on the average of the 6 values obtained from the two sessions.

Both Intraclass Correlation Coeffecient (ICC) and Standard Error of Measurement (SEM) were used to characterize intra- and intersession repeatability. The ICC method and nomenclature were those proposed in (McGraw and Wong, 1996). The ICC(A,1) model was used, which assesses the degree of absolute agreement among measurements. The SEM is the group-level estimation of the intra-subject average variability (Weir, 2005), defined by the following equation: $SEM = S_T \sqrt{1-R}$, where $S_T$ is the global standard deviation and $R$ is the corresponding ICC. From SEM, we also computed the Smallest Detectable Difference ($SDD = SEM \ x \ 1.96 \ x \ \sqrt{2}$), which is the smallest change that could be considered as significant (Weir, 2005). Normalized by the mean and expressed as percentage, it allows to evaluate whether an observed relative change in one individual could be true and not due to measurement error or to intra-individual variability.

The reliability of the average of 2 repetitions of 35 strides was assessed by using the Spearman-Brown prophecy formula, which predicts the reliability composed by replicating a given test:

$$R^* = \frac{NR}{1 + (N-1)R}$$

where $R^*$ is the predicted reliability, $N$ is the number of tests (in this case N=2) and $R$ is the measured reliability for one test. The purpose was to test whether the predicted reliability for the average of 2x35 strides corresponded to the measured reliability obtained from 70 strides. A substantial difference between predicted and measured reliability would indicate that the error strongly depends upon the length of the measurement.

Practically, the reliability within the two sessions (intrasession repeatability, based on three repetitions, ICC(A,1), SEM and SDD) was assessed separately for each session, and then the results were averaged (table 1). Two complementary approaches were used to characterize the week-to-week reliability: first, we assessed the 70-strides intersession repeatability (two repetitions) by computing pairwisely three ICCs obtained from each 70-strides measurements taken from the two sessions, and then averaging them (table 2). Second, the LDS results of the three 70-strides measurements were averaged (3x70 strides = 210), and the repeatability between the sessions (two repetitions) was computed (table 3). The first approach provided information on the reliability of short duration walking test (70 strides), while the second approach mimics the reliability of discontinuous walking tests, from which three repetitions are used to compute an average result (210 strides). The confidence intervals on the estimates (95% CI) were computed by bootstrapping (5000 resamples, bias corrected and accelerated percentile method). A graphical representation of the reliability results is provided in the online supplemental material.

## 3. Results

Regarding the descriptive statistics of LDS (fig. 2), the results are normally distributed among individuals (Lilliefors test, p>0.05), with some outliers. A substantial difference exists between the estimates obtained from 35 strides and from 70 strides, with a more marked effect in long-term LDS; the relative differences (70 strides - 35 strides / 35 strides x 100) were on average +40% for $\lambda_{4\text{-}10}$, +6% for $\lambda_1$, and +8% for $\lambda_{0.5}$.

The results of the intra-session repeatability (ICC and SEM) are shown in the table 1. Regarding the 35-strides estimates, a substantial difference exists between long-term LDS (ICC 0.17-0.20) and short-term LDS (ICC 0.71-0.82). Namely, the 70-strides predicted repeatability (column #7) is comparable to the measured repeatability for short-term LDS, but not for long-term LDS. The 70-strides LDS estimates exhibit higher reliability, with the best relative SDDs for $\lambda_{0.5}$ (11%-13%).

The results of the intersession repeatability are presented in table 2 and 3. Based on 70-strides estimates, the reliability is very low for long term-LDS (SDD 102%-142%) and is maximal for $\lambda_{0.5}$ (SDD 19%-25%). Based on 210 strides estimates, the ICCs are homogenous ($\lambda_{4\text{-}10}$: 0.58, $\lambda_1$: 0.59, $\lambda_{0.5}$: 0.60). In contrast, the SEM and SDD estimates are higher for long-term LDS, and minimal for $\lambda_{0.5}$ (SDD 16%-23%). The medio-lateral direction tends to exhibit lower SEM and SDD.

## 4. Discussion

Referring to the research questions presented in the introduction, the results can be summarized as follows: 1) the intrasession repeatability of gait LDS (ICC, 70-strides estimates) was around 0.50 for long-term LDS and 0.85 for short-term LDS, the intersession repeatability was around 0.6 for both short- and long-term LDS estimated from 210 strides (3x70). 2) Long-term LDS estimated from short-duration measurements (35 strides) exhibited particularly low repeatability (ICC: 0.20), while that of short-term LDS was around 0.75. 3) Long-term LDS exhibited a substantial discrepancy between the predicted repeatability and the actual repeatability, while it was not the case for short-term LDS. 4) Taking into account both ICC and SEM, short-term LDS measured over one step ($\lambda_{0.5}$) tended to have better intra- and inter-session reliability than short-term LDS measured over one stride ($\lambda_1$).

LDS has been computed from various kinematic signals (speed, acceleration, joint angles) measured at different level (lower limbs, trunk) with different measurement methods (video analysis, accelerometers). The present study used trunk acceleration measured at the anterior part of the thorax (sternum), a method that has been originally proposed in the first studies that introduced Lyapunov exponents for gait stability assessment (Dingwell and Cusumano, 2000; Dingwell et al., 2001). Few studies have been conducted to compare the validity and reliability among measurement methods (Bruijn et al., 2010b; Kang and Dingwell, 2009). Based on theoretical considerations and empirical evidence, it is thought that different methods can be alternatively used to record the gait dynamics with similar results (Bruijn et al., 2013). The results of the present study are likely generalizable to other measurement methods, but further reliability studies are needed to confirm that hypothesis.

The results of the present study clearly show that $\lambda_{0.5}$ is more reliable than $\lambda_1$. Indeed, regarding the intrasession reliability, the $\lambda_{0.5}$ repeatability (ICC) is on average 6% higher than the $\lambda_1$ repeatability. Furthermore, the relative SDD was on average 12% for $\lambda_{0.5}$ and 20% for $\lambda_1$. Regarding intersession reliability computed on 210 strides (table 3), although both $\lambda_{0.5}$ and $\lambda_1$ exhibit similar ICC (0.60 vs. 0.59), the relative SDD is substantially lower for $\lambda_{0.5}$ (average SDD: 20% vs. 28%). The same conclusion can be drawn from the analysis of the 70-strides results (table 2). Although further investigations are needed to compare the responsiveness and validity of both $\lambda_{0.5}$ and $\lambda_1$, it is recommended to use $\lambda_{0.5}$ in order to increase the statistical power (Perkins et al., 2000).

We confirm that the estimates of divergence exponents ($\lambda$) increases with the measurement length (Bruijn et al., 2009; Kang and Dingwell, 2006). The relative effect for doubling the length of measurement (35 strides to 70 strides) was on average +40% for long-term LDS and +7% for short-term LDS. An explanation could be that a long-duration signal increases the probability of occurrence of very close neighbors that have a larger potential of divergence away from each other (Bruijn et al., 2009). As a result, a careful normalization of sample lengths is required when comparing LDS among groups and conditions, especially if long-term LDS is analyzed.

Since the classical works of Spearman (Spearman, 1910) and Brown (Brown, 1910), it is well known that the mean of several replicate measurements on a subject is more reliable than a single measurement. Accordingly, the short-term LDS estimated from 70-strides exhibited repeatability, which was similar to repeatability of the average of two 35-strides measurements such as predicted by the Spearman-Brown formula (table 1). Therefore, there is no fundamental flaw in the Rosenstein's algorithm that would hinder the computation of the short-term divergence over a small number of consecutive strides. Although short-term LDS may substantially vary from one stride to another, the reliability can be enhanced by averaging several repetitions of short overground walking trials, as already observed in another study (van Schooten et al., 2011). That opens the perspective of LDS assessment in patients with diminished walking capabilities, with sufficient time between successive repetitions to offer recuperation. However, that relies upon the assumption that short repeated overground walking tests will exhibit the same reliability as the continuous treadmill walking, which we used in the present study. We recently obtained results in overground walking that confirm a good intrasession reliability of two consecutive repetitions of 40 strides (ICC about 0.85)(Terrier and Reynard, 2013). However, more studies are needed to assess the intra-day reliability of LDS. In contrast to short-term LDS, long-term LDS exhibited very low repeatability when estimated with 35 strides only (table 1). The increase in reliability using 70-strides estimates was substantially larger than predicted. Thus, the Rosenstein algorithm needs a continuous signal of substantial length to precisely assess divergence over longer time scale.

While there is likely a consensus to judge an ICC value of 0.3 as "low" and an ICC value of 0.95 as "high", there is no absolute threshold to delimit whether a reliability index is "acceptable" (Lance et al., 2006). On the contrary, the satisfactory reliability level depends upon the use of the investigated variable. In particular, the responsiveness (effect size) of the variable should guide the interpretation of reliability results. Various studies have demonstrated a large spread in LDS sensitivity to different conditions, as for example: when submitting individuals to mechanical perturbations during treadmill walking, the short-term LDS is lowered by about 70% (McAndrew et al., 2011); fall-prone older adults exhibits a lower LDS as compared to healthy controls: the difference is -20% (Lockhart and Liu, 2008); when balance is artificially impaired by using galvanic vestibular stimulation (GVS), the mean relative effect on short-term LDS is -11% (van Schooten et al., 2011); the use of orthopedic shoes tends to improve short-term LDS in patients with chronic foot & ankle injuries (+10% in ML direction (Terrier et al., 2013)). In experimental studies, which submit individuals to varying conditions during a single testing session, the short-term LDS reliability (table 1) is likely sufficient to highlight gait stability modification even when using 35-strides estimates, at least at group level. However, such a study design is of limited relevance for clinical studies, which most often aim to assess long term therapy efficiency in longitudinal studies (Yakhdani et al., 2010). In particular, the use of LDS as a diagnostic tool for fall risk assessment at individual level needs a sufficiently low intersession SEM and SDD (table 2 and 3). In this context, only $\lambda_{0.5}$ measured in ML direction averaged from 210 strides seems reliable enough (SDD: 16%) to highlight a change that corresponds to the difference between healthy and fall-prone older adults (20% (Lockhart and Liu, 2008)). Therefore, there is still a need to further analyze the origin of the substantial week-to-week variability of LDS. As explanation, one could postulate a modification of sensor placement. Other causes could be change of individual's mood and fatigue.

Interestingly, it seems that a difference exists between the reliability of kinematic parameters and the reliability of variability indices. Indeed, high ICC (>0.90) have been reported for both temporal and spatial kinematic parameters measured in short walking tests either for intersession reliability (van Uden and Besser, 2004) or intrasession reliability (Brach et al., 2008). In contrast, the classical measures of variability exhibits lower repeatability: ICCs between 0.40 and 0.63 have been reported for intrasession repeatability (SD of step length and time) (Brach et al., 2008). In the same manner, it has been reported that a substantial number of consecutive strides (>60) are needed to precisely assess the variability of gait velocity (Hollman et al., 2010). Therefore, LDS estimates seem to be sufficiently precise as compared as other (classical) gait variability indices.

## 5. Conclusions

To conclude, the following advices should be given for future gait LDS studies: 1) It is recommended to normalize sample length before computing LDS to thwart the trend to higher LDS estimates with longer measurements. 2) Regarding long-term LDS, its low reliability when few strides are analyzed make the use of a treadmill highly recommended in order to record long duration walking tests. Furthermore, given the limited reliability, its use should be restricted to group-level assessment with a sufficient sample size to lower type II error risk. 3) Regarding short-term LDS, it is recommended to use $\lambda_{0.5}$, provided that future studies further document its validity. 4) The ML acceleration signal should be used in priority because it exhibits the lowest SEM. 5) As far as treadmill results could be applied to overground situations, the repetition of several short-duration indoor walking tests would be a solution to assess gait stability in clinical settings. 6) The reliability of short-term LDS may be sufficient to detect moderate changes (around 20%) at individual level. However, due to the lower intersession repeatability, one should aggregate several measurements taken on different days, in order to better approximate the true LDS.

### Acknowledgments

The authors would like to thank Olivier Dériaz for his valuable support and thoughtful advice. The study was supported by the Swiss accident insurance company SUVA, which is an independent, non-profit company under public law, and by the clinique romande de réadaptation. The IRR (Institute for Research in Rehabilitation) is supported by the State of Valais and the City of Sion.

**Conflict of interest statement**

There are no known conflicts of interest.

| | | 35 Strides | | | | 70 str. | 70 Strides | | | | Relative SDD | |
|---|---|---|---|---|---|---|---|---|---|---|---|---|
| | | ICC | 95% CI | SEM | 95% CI | predicted | ICC | 95% CI | SEM | 95% CI | 35 str. | 70 str. |
| **Long-term LDS ($\lambda_{4\text{-}10}$)** | AP | **0.20** | 0.11 - 0.30 | 0.010 | 0.009 - 0.011 | 0.33 | **0.54** | 0.46 - 0.63 | 0.006 | 0.006 - 0.007 | **275%** | **118%** |
| | V | **0.17** | 0.07 - 0.30 | 0.014 | 0.013 - 0.015 | 0.29 | **0.54** | 0.45 - 0.64 | 0.008 | 0.008 - 0.009 | **231%** | **95%** |
| | ML | **0.18** | 0.08 - 0.30 | 0.011 | 0.010 - 0.012 | 0.31 | **0.48** | 0.39 - 0.59 | 0.007 | 0.006 - 0.008 | **233%** | **106%** |
| **Short-term LDS ($\lambda_1$)** | AP | **0.78** | 0.73 - 0.84 | 0.041 | 0.037 - 0.045 | 0.87 | **0.85** | 0.81 - 0.89 | 0.034 | 0.031 - 0.038 | **24%** | **19%** |
| | V | **0.74** | 0.67 - 0.80 | 0.058 | 0.053 - 0.065 | 0.85 | **0.83** | 0.79 - 0.88 | 0.047 | 0.043 - 0.053 | **27%** | **21%** |
| | ML | **0.71** | 0.65 - 0.76 | 0.036 | 0.034 - 0.040 | 0.83 | **0.79** | 0.75 - 0.84 | 0.030 | 0.028 - 0.033 | **24%** | **19%** |
| **Short-term LDS ($\lambda_{0.5}$)** | AP | **0.82** | 0.79 - 0.86 | 0.060 | 0.055 - 0.065 | 0.90 | **0.88** | 0.86 - 0.91 | 0.049 | 0.045 - 0.055 | **15%** | **12%** |
| | V | **0.82** | 0.77 - 0.87 | 0.073 | 0.067 - 0.081 | 0.90 | **0.88** | 0.85 - 0.91 | 0.061 | 0.055 - 0.069 | **17%** | **13%** |
| | ML | **0.74** | 0.69 - 0.79 | 0.044 | 0.040 - 0.048 | 0.85 | **0.81** | 0.78 - 0.85 | 0.037 | 0.034 - 0.039 | **15%** | **11%** |

***Table 1. Intrasession reliability of local dynamic stability.*** *Ninety-five participants walked during five minutes on a treadmill at two occasions separated by a week. The local dynamic stability (LDS) was computed using the maximal finite-time Lyapunov exponents method from trunk acceleration signal measured in antero-posterior (AP), vertical (V), and medio-lateral (ML) directions. Three time scales were used: long-term LDS between 4 and 10 strides ($\lambda_{4\text{-}10}$); short-term LDS between 0 and 1 stride ($\lambda_1$); short-term LDS between 0 and 0.5 stride ($\lambda_{0.5}$). The LDS estimates were computed either from 35 strides (left) or from 70 strides (right). The agreement among three within-session repetitions (intraclass correlation coefficient (ICC(A,1)), the standard error of measurement (SEM), and the relative smallest detectable difference (SDD) are shown. The repeatability for a 70 strides test predicted from a 35 strides test (Spearmann-Brown) is also shown. The ICCs, SEMs and SDDs obtained during both sessions were averaged. The 95% confidence intervals (CI) were computed by bootstrapping (5000 resamples ).*

| | | ICC | 95% CI | | SEM | 95% CI | | Relative SDD |
|---|---|---|---|---|---|---|---|---|
| **Long-term LDS** ($\lambda_{4\text{-}10}$) | **AP** | **0.35** | 0.24 - 0.47 | | 0.007 | 0.007 - 0.008 | | **142%** |
| | **V** | **0.47** | 0.36 - 0.59 | | 0.009 | 0.008 - 0.010 | | **102%** |
| | **ML** | **0.37** | 0.26 - 0.51 | | 0.008 | 0.007 - 0.009 | | **117%** |
| **Short-term LDS** ($\lambda_1$) | **AP** | **0.46** | 0.32 - 0.58 | | 0.064 | 0.057 - 0.073 | | **36%** |
| | **V** | **0.60** | 0.47 - 0.71 | | 0.073 | 0.065 - 0.084 | | **32%** |
| | **ML** | **0.53** | 0.42 - 0.62 | | 0.046 | 0.042 - 0.054 | | **29%** |
| **Short-term LDS** ($\lambda_{0.5}$) | **AP** | **0.51** | 0.37 - 0.63 | | 0.102 | 0.091 - 0.117 | | **25%** |
| | **V** | **0.63** | 0.53 - 0.75 | | 0.106 | 0.095 - 0.122 | | **23%** |
| | **ML** | **0.50** | 0.40 - 0.61 | | 0.060 | 0.054 - 0.067 | | **19%** |

**Table 2. Intersession (week-to-week) reliability of local dynamic stability based on 70 strides.** *Ninety-five participants walked during five minutes on a treadmill at two occasions separated by a week. The local dynamic stability (LDS) was computed using the maximal finite-time Lyapunov exponents method from trunk acceleration signal measured in antero-posterior (AP), vertical (V), and medio-lateral (ML) directions. Three time scales were used: long-term LDS between 4 and 10 strides ($\lambda_{4\text{-}10}$); short-term LDS between 0 and 1 stride ($\lambda_1$); short-term LDS between 0 and 0.5 stride ($\lambda_{0.5}$). The agreement among two between-session repetitions of 70-strides length (intraclass correlation coefficient (ICC(A,1)), the standard error of measurement (SEM), and the relative smallest detectable difference (SDD) are shown. The 95% confidence intervals (CI) were computed by bootstrapping (5000 resamples).*

| | | ICC | 95% CI | | SEM | 95% CI | | Relative SDD |
|---|---|---|---|---|---|---|---|---|
| **Long-term LDS** ($\lambda_{4\text{-}10}$) | AP | **0.47** | 0.31 - 0.61 | | 0.0055 | 0.005 - 0.007 | | **107%** |
| | V | **0.67** | 0.57 - 0.77 | | 0.0058 | 0.005 - 0.006 | | **67%** |
| | ML | **0.60** | 0.47 - 0.73 | | 0.0050 | 0.004 - 0.006 | | **76%** |
| **Short-term LDS** ($\lambda_1$) | AP | **0.51** | 0.36 - 0.63 | | 0.059 | 0.051 - 0.068 | | **33%** |
| | V | **0.66** | 0.52 - 0.78 | | 0.064 | 0.055 - 0.076 | | **28%** |
| | ML | **0.60** | 0.48 - 0.70 | | 0.040 | 0.035 - 0.047 | | **25%** |
| **Short-term LDS** ($\lambda_{0.5}$) | AP | **0.55** | 0.40 - 0.67 | | 0.094 | 0.083 - 0.110 | | **23%** |
| | V | **0.68** | 0.57 - 0.80 | | 0.096 | 0.084 - 0.113 | | **21%** |
| | ML | **0.58** | 0.47 - 0.68 | | 0.051 | 0.045 - 0.060 | | **16%** |

***Table 3. Intersession (week-to-week) reliability of local dynamic stability based on 210 strides.*** *Ninety-five participants walked during five minutes on a treadmill at two occasions separated by a week. The local dynamic stability (LDS) was computed using the maximal finite-time Lyapunov exponents method from trunk acceleration signal measured in antero-posterior (AP), vertical (V), and medio-lateral (ML) directions. Three time scales were used: long-term LDS between 4 and 10 strides ($\lambda_{4\text{-}10}$); short-term LDS between 0 and 1 stride ($\lambda_1$); short-term LDS between 0 and 0.5 stride ($\lambda_{0.5}$) The LDS estimates were computed from 210 strides (average of the three 70 strides estimates). The agreement among two repetitions (intraclass correlation coefficient (ICC(A,1)), the standard error of measurement (SEM), and the relative smallest detectable difference (SDD) are shown. The 95% confidence intervals (CI) were computed by bootstrapping (5000 resamples).*

# Figure captions

**Figure 1. Overall view of the experimental procedure and of the data pre-processing.** *LDS: local dynamic stability. PWS: preferred walking speed. AP: antero-posterior. V: Vertical. ML: medio-lateral.*

**Figure 2. Descriptive statistics of the local dynamic stability (LDS).** *Ninety-five participants walked during five minutes on a treadmill at two occasions separated by a week. LDS was computed using the maximal finite-time Lyapunov exponents method from trunk acceleration signal measured in antero-posterior (AP), vertical (V), and medio-lateral (ML) directions. Three time scales were used: long-term LDS between 4 and 10 strides ($\lambda_{4\text{-}10}$); short-term LDS between 0 and 1 stride ($\lambda_1$); short-term LDS between 0 and 0.5 stride ($\lambda_{0.5}$). The LDS estimates were computed either from 35 strides (black) or from 70 strides (grey). The results of both sessions were averaged (see fig. 1). The spread of the data among participants is presented with boxplots (median and quartiles) and with 95% confidence intervals around the mean (vertical lines). + indicates outliers. Printed values are mean and SD (N=95).*

**Figure 1**

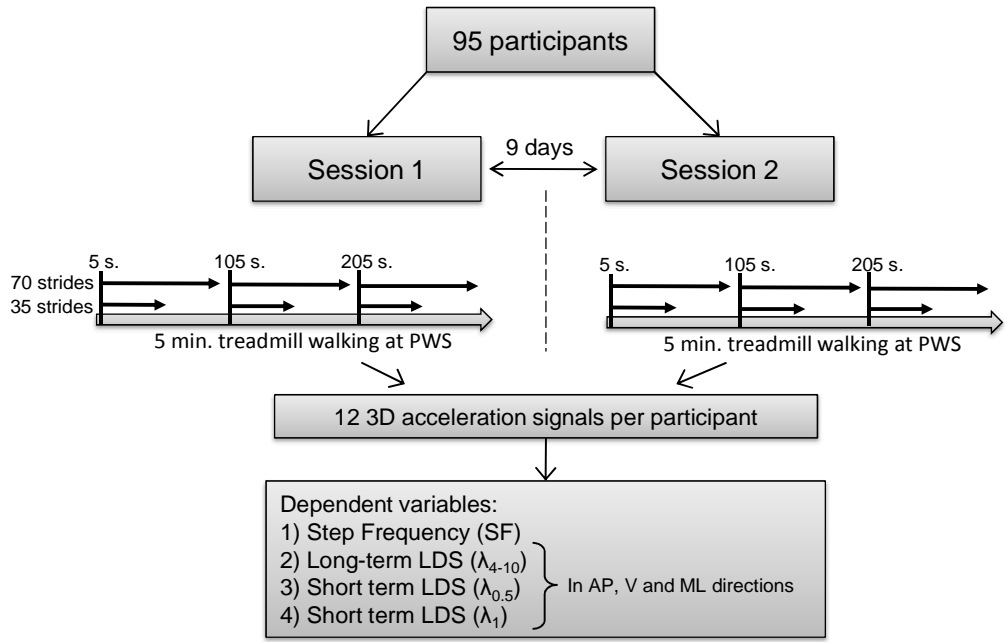

**Figure 2**

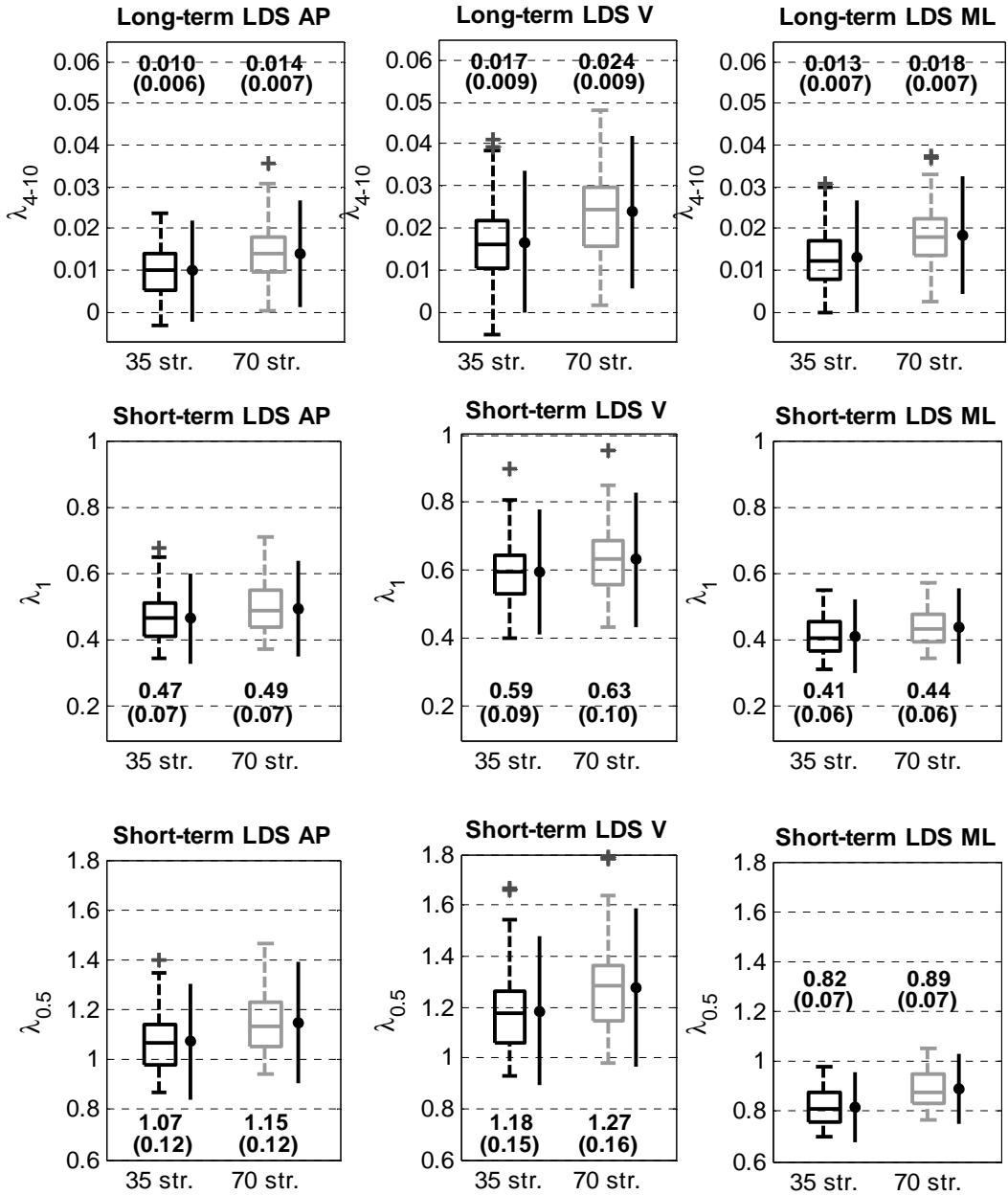